\newif\ifarxiv
\arxivtrue

\ifarxiv
  \documentclass[11pt]{article}
\else
  \documentclass[mnsc,sglanonrev]{informs4}
  \usepackage{eqndefns-left} 
\fi

\RequirePackage{tgtermes}
\RequirePackage{newtxtext}
\RequirePackage{newtxmath}
\RequirePackage{bm}
\RequirePackage{endnotes}

\ifarxiv\else\fi

\ifarxiv\else
  \OneAndAHalfSpacedXII 
\fi

\usepackage{amsmath}
\usepackage{graphicx}
\usepackage{booktabs}
\usepackage{array}
\usepackage{multicol}
\usepackage{enumitem}
\usepackage{verbatim}
\usepackage{xcolor}

\usepackage{hyperref}
\usepackage{etoolbox}
\usepackage{setspace}

\definecolor{sciblue}{RGB}{0,84,147}

\newcolumntype{L}[1]{>{\raggedright\arraybackslash}p{#1}}
\newcolumntype{C}[1]{>{\centering\arraybackslash}p{#1}}

\hypersetup{colorlinks,
citecolor=blue,
linkcolor=blue,
urlcolor=blue}

\usepackage{natbib}

 \bibpunct[, ]{(}{)}{,}{a}{}{,}%
 \def\bibfont{\small}%
 \def\bibsep{\smallskipamount}%

\ifarxiv
  \usepackage[margin=1in]{geometry}
  \usepackage[misc]{ifsym} 
  \newcommand{\RUNAUTHOR}[1]{}
  \newcommand{\RUNTITLE}[1]{}
  \newcommand{\TITLE}[1]{\title{#1}}
  \newcommand{\savedabstract}{}
  \newcommand{\ABSTRACT}[1]{\renewcommand{\savedabstract}{#1}}
  \newcommand{\savedkeywords}{}
  \newcommand{\KEYWORDS}[1]{\renewcommand{\savedkeywords}{#1}}
  \newcommand{\ARTICLEAUTHORS}[1]{\author{#1}}
  \newcommand{\AUTHOR}[1]{#1}
  \newcommand{\AFF}[1]{\thanks{#1}}
  \newcommand{\EMAIL}[1]{\Letter\ #1}
  \let\informsmaketitle\maketitle
  \renewcommand{\maketitle}{%
    \informsmaketitle
    \begin{abstract}
    \savedabstract
    \end{abstract}
    \noindent\textbf{Keywords:} \savedkeywords\par\bigskip
  }
  \newcommand{\TABLE}[3]{\caption{#1}#2\par\smallskip\noindent{\footnotesize #3}}
  \newcommand{\FIGURE}[3]{\centering#1\caption{#2}\par\smallskip\noindent{\footnotesize #3}}
  \newenvironment{APPENDICES}{\appendix}{}
  \newcommand{\ACKNOWLEDGMENT}[1]{\par\bigskip\noindent\textbf{Acknowledgments.} #1}
\else
  \EquationsNumberedThrough    

  \TheoremsNumberedThrough     
  \ECRepeatTheorems  %
  \MANUSCRIPTNO{}

  \def\ABSfont{\fontsize{10.8}{16}\selectfont}
\fi
\begin{document}

\RUNAUTHOR{Coqueret et al.}

\RUNTITLE{Randomness in large language models}

\TITLE{Randomness in large language models:\\ What researchers need to know (and report)}

\ifarxiv
\ARTICLEAUTHORS{%
\AUTHOR{Guillaume Coqueret}%
\AFF{EMLYON Business School, France, \EMAIL{coqueret@em-lyon.com}. Coqueret is affiliated with \href{https://cascad.tech/}{Cascad}, a nonprofit certification agency that verifies the reproducibility of research articles in economics and management.}
\and
\AUTHOR{Joan Llull}%
\AFF{Institute for Economic Analysis (IAE-CSIC), and Barcelona School of Economics, Spain, \EMAIL{joan.llull@iae.csic.es}. Llull is the \href{https://www.econometricsociety.org/publications/es-data-editor-website}{Data Editor of the Econometric Society}. Llull acknowledges funding from  the European Research Council (ERC), through Consolidator Grant n.~10112542, the Spanish Ministry of Economy and Competitiveness, through grant PID2024-162507OB-I00 and from the Severo Ochoa Programme for Centres of Excellence in R\&D (Barcelona School of Economics CEX2024-001476-S), and by the Generalitat de Catalunya (2025-SGR-00973).}
\and
\AUTHOR{Florian Oswald}%
\AFF{University of Turin and Collegio Carlo Alberto, Italy, \EMAIL{florian.oswald@unito.it}. Oswald is the \href{https://jpedataeditor.github.io}{Data Editor of the Journal of Political Economy}.}
\and
\AUTHOR{Christophe P\'erignon}%
\AFF{HEC Paris, France, \EMAIL{perignon@hec.fr}. P\'erignon is co-founder and Director of \href{https://cascad.tech/}{Cascad}.}
\and
\AUTHOR{Christoph Scheuch}%
\AFF{Humboldt University of Berlin, Germany, \EMAIL{christoph.scheuch@hu-berlin.de}. Scheuch is the \href{https://review-of-financial-studies.github.io}{Data Editor of the Review of Financial Studies} and is affiliated with \href{https://cascad.tech/}{Cascad}.}
\and
\AUTHOR{Lars Vilhuber}%
\AFF{Cornell University, United States, \EMAIL{lars.vilhuber@cornell.edu}. Vilhuber is the \href{https://aeadataeditor.github.io/}{Data Editor of the American Economic Association}.}
}
\fi

\ABSTRACT{%
Large language models (LLMs) are increasingly used to generate data for research. Typical use cases are classifications, annotations, information extraction, and generation of numerical scores. Unlike conventional measurements, LLM outputs can vary across repeated requests even when the prompt and apparent model settings remain unchanged. This variation arises from deliberate sampling, silent model updates, numerical rounding, or expert routing. Setting a dedicated \textit{temperature} parameter to zero, when that option is available, targets deliberate sampling but does not address the other sources of randomness. Exact regeneration of outputs is therefore generally not possible when using proprietary application programming interfaces. Local execution of open-weight models offers greater control, but exact regeneration still depends on the complete hardware and software stack. We illustrate these issues through sentiment classifications of corporate filings and examine their consequences for downstream regression results. We then propose practical guidance on what authors should report about the variability of LLM-generated data, what editors and reviewers should assess, and what data editors should verify.
}%



\maketitle


\section{Introduction}\label{sec:Intro}

Large language models (LLMs) have become ubiquitous in modern academic research. Statements such as "\textit{we use a large language model to classify 60,000 documents}" now appear routinely in empirical studies in economics and management. Researchers use these models to classify and annotate corporate filings \citep{wang2025assessing} and earnings announcements \citep{siano2025news,harford2026}, measure sentiment or identify topics in social media posts and news articles \citep{gilardi2023}, interpret and summarize central bank communications \citep{hansen2023}, extract structured information from historical records and images \citep{dell2025}, predict financial outcomes \citep{lopezlira2025}, and generate customer recommendations \citep{kweon2025uncertainty}. The resulting model outputs increasingly serve as inputs for subsequent analyses.

Unlike traditional measurements, LLM-generated data are generally \textit{not deterministic} \citep{ouyang2024,he2025,deKokMS2025}. The same prompt may yield different outputs, and therefore different data, even when the prompt and apparent model settings are unchanged. For commercial providers such as OpenAI or Anthropic, researchers typically cannot observe the serving infrastructure, archive the deployed model, identify the hardware on which it ran, or even control the \textit{temperature} (a parameter akin to a logit scale, determining the amount of randomness in model output) in every interface. These hidden sources of variation are rarely acknowledged, and the resulting data are typically analyzed as if they were fixed observations.

Existing data and code policies adopted by academic journals in the social sciences, like economics and management, offer only limited guidance for this situation. Most reproducibility standards require authors to document sources of randomness and provide code that controls them whenever possible \citep{vilhuber2019,koren2022,vilhuber2022}. These standards, however, largely predate the widespread adoption of generative AI and provide little guidance on how to report data generated by probabilistic black-box models. This lack of disclosure and guidance has important implications for the transparency and reproducibility of research, which are central concerns for academic journals \citep{fivsar2024reproducibility,loch2025reinforcing}.

Our central thesis is simple: data generated by an LLM should be treated as draws from a distribution, not as fixed measurements. To develop this perspective, we make four contributions. First, we identify the main sources of variation across repeated model calls and explain the underlying mechanisms in a way that is accessible to the broad community of empirical researchers. Second, we distinguish sources of variation that researchers can control from those that remain under the control of model providers. Third, we propose an experiment that measures the consequences of these sources of variation in a typical task involving sentiment annotation of annual corporate filings. Fourth, and most important, we translate these points into practical guidance on what authors should report, what editors and reviewers should assess, and what data editors should verify. We then map this guidance against the current \textit{Management Science} policy, as well as other policies and standards. We hope these contributions will stimulate discussion among researchers and journal (data) editors, and provide a foundation for the development of reporting and verification standards for the use of LLMs in empirical research.

We complement the emerging literature on the reproducibility and reliability of LLM-generated data in empirical research. \citet{kim2026} shows that reliance on a single model run can produce unstable coefficients and misleading significance. The appropriate solution depends on whether model output is itself the research outcome or is an input for subsequent analysis. Related work by \citet{wang2025assessing} examines the consistency and reproducibility of LLM outputs across a range of finance and accounting tasks. They show that reproducibility varies across tasks, while simple aggregation over multiple model runs can substantially improve consistency. \citet{deKokMS2025} provides practical guidance for using LLMs in empirical research and is among the first papers to discuss the reproducibility implications of model-generated data. The paper recommends archiving prompts and raw model outputs, noting that commercial models may change or become unavailable over time.

\section{How researchers use generative artificial intelligence}\label{sec:researcher_use}

To document the rapid adoption of LLMs in economics and management research, we examine articles published between January 2020 and November 2026 in the 145 top-tier academic journals ranked 4 or 4* in the \href{https://charteredabs.org/academic-journal-guide}{Academic Journal Guide} (out of the 1,822 journals listed in the Guide). To identify articles using LLMs, we search the abstracts indexed by \href{https://www.openalex.org}{OpenAlex} for mentions of 212 LLMs
, as well as generic terms such as "LLM", "large language model", "Gen AI", "generative AI", and "generative artificial intelligence".

\begin{figure}
\FIGURE
{\includegraphics[width=0.8\textwidth]{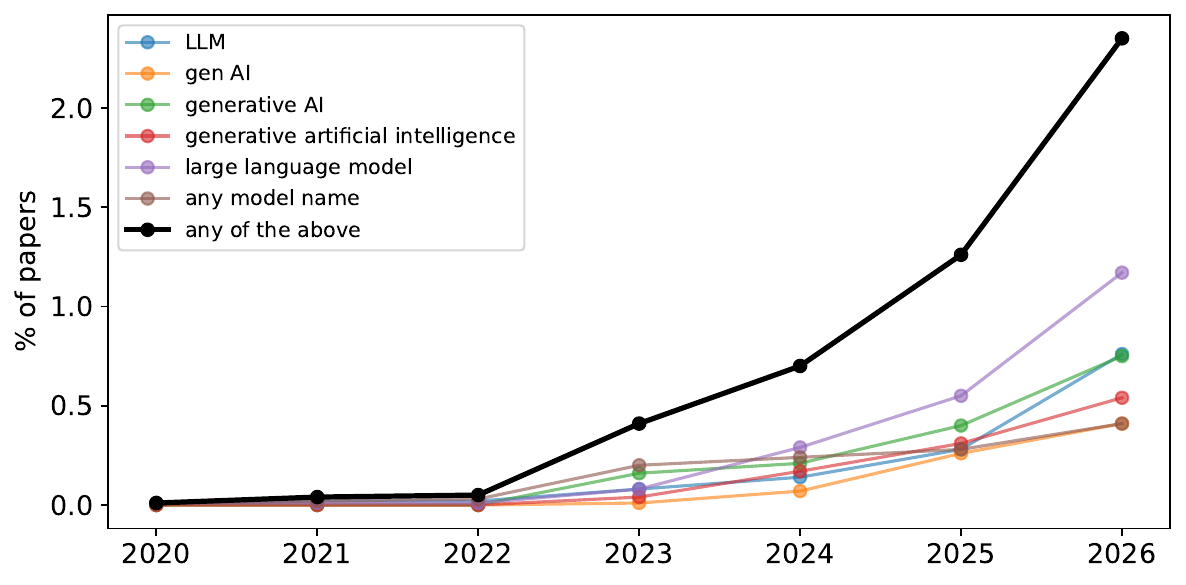}}
{Use of LLMs in economics and business journals.\label{fig:term_freq}}
{We plot the share of articles whose titles or abstracts feature a set of keywords associated with the use of LLMs. The sample consists of the 145 journals ranked 4 or 4* in the Academic Journal Guide ranking list. "Any model name" means that papers are flagged whenever they mention a specific LLM in their title or in their abstract. The total annual number of articles published across the 145 journals are 15,545 (2020), 15,615 (2021), 14,573 (2022), 13,700 (2023), 14,185 (2024), 14,664 (2025), and 14,312 (2026).}
\end{figure}

Figure \ref{fig:term_freq} shows the share of articles in our sample of 101,892 publications that mention the above keywords. The increase is striking for all keyword groups. As of November 2026, roughly 2.4\% of papers published in leading economics and management journals mention LLMs in either their title or abstract. Although already substantial, this estimate is likely conservative because it is based solely on titles and abstracts rather than the full text of articles. Moreover, published articles reflect research practices only with a lag of one to three years because of the time required for peer review and publication.\footnote{Our count shows how often articles mention LLMs, not how they report their use. A full-text survey could show how often authors omit the model version, query date, or temperature, which we leave to future work.} Consistent with this view, a \href{https://www.linkedin.com/feed/update/urn:li:activity:7486740249776242688/}{July 2026 NBER survey of economists} found that 60\% reported LLMs to be an essential part of their research workflow.

The increasing prevalence of LLMs documented in Figure \ref{fig:term_freq} raises a natural question: how are these models actually used in research in economics and management? We classify these uses into three main categories. The first is data generation and measurement. Researchers use models to classify text, annotate images, extract quantities, code qualitative evidence, and assign numerical scores. The outputs may become outcomes, explanatory variables, controls, or sample selection rules. This use creates the most direct reproducibility risk because LLM output becomes part of the data used in the analysis.

The second category involves simulated respondents and agents. Models can serve as substitutes for survey respondents or as agents in economic environments \citep{horton2023}. This approach can generate useful hypotheses, but it also raises questions about representativeness, embedded social biases, and the validity of replacing human behavior with model behavior \citep{bail2024}. The third category is research assistance, including programming, literature searches, writing, translation, and experimental design \citep{charness2023,korinek2023,korinek2025}. These uses create important disclosure and verification questions, but they usually leave artifacts that researchers can inspect. A human can review generated code, confirm a citation, or revise generated prose.

We focus on the first category because it has the most direct implications for research reproducibility. The central case is one in which model output becomes data. If a model determines a sentiment variable, identifies observations for inclusion, or measures an outcome, variation in its output can change the published result. Reproducibility therefore depends on the model and its serving environment as well as on the conventional data and code package.

Researchers' typical use of an LLM is as follows. Through an \textit{application programming interface} (API), a program sends a request over the internet and a provider runs the model on its own servers. \textit{Open weights} are a model's published numerical parameters, which let anyone with suitable hardware run the model locally. \textit{Closed-weight} models are available only through their provider. Requests often include a \textit{system message}, a set of standing instructions sent alongside the user's prompt. The \textit{serving environment} is everything that executes the model: hardware, drivers, inference software, numerical precision, and the batching of requests.


\section{Where the randomness comes from}\label{sec:randomness}

A language model produces text one token at a time. A token can be a word, part of a word, a number, or another unit used by the model. At each step, the model assigns a so-called logit score to every candidate token. These logits are transformed into probabilities or used directly to select the next token. Given a certain \emph{context} (a set of preceding tokens), the model produces a probability distribution over potential next tokens -- in practice, a vector of length $|V|$, which stands for the size of the model's vocabulary $V$, i.e. all tokens known to the model.

Consider the incomplete sentence "\textit{John Maynard Keynes lived}". Suppose the LLM assigns probability 0.70 to token "\textit{from}", probability 0.15 to "\textit{in}", probability 0.10 to "\textit{during}", and probability 0.05 to all other candidates combined. LLMs typically apply a sampling procedure over those potential candidates, so as to make the model output less dull and repetitive. A greedy procedure would select "\textit{from}" because it has the highest probability. The selected token is then added to the context, and the model calculates a new set of logits for the next token.

This sequential structure matters. A small difference at one step can alter every later step because later logits are conditional on the text already produced. Minor numerical variation can therefore lead to substantively different final answers.

\subsection{Deliberate sampling}

The most visible source of variation is deliberate sampling. Under sampling, the model draws the next token from the generated distribution over $V$ rather than greedily choosing the token with the highest logit score. This design makes conversational output more varied and natural. The distribution over next tokens $x$ is generated as follows, whereby where $z_i$ is the logit of token $i$, $|V|$ is the total number of possible tokens a model can choose from, and $T$ is the temperature:
\begin{equation}
\label{eq:softmax}
P(x_i \mid \text{context}) \;=\; \frac{\exp(z_i/T)}{\sum_{j=1}^{|V|} \exp(z_j/T)},
\qquad T > 0.
\end{equation}
A high temperature flattens the distribution toward uniform, so that all tokens have similar probabilities of being chosen. A low temperature narrows it around the largest logit. In the limit, when the temperature is equal to zero (\textit{greedy decoding}), Equation \eqref{eq:softmax} is not defined and the model chooses the highest-scoring token at each step. In the case of a tie, the token with the smallest token index $j$ is usually chosen.\footnote{$T=0$ does not always mean greedy decoding. Some inference software switches to the highest-scoring token \citep{vllm2026,mlxvlm2026}, other software rejects zero, so users substitute a small positive value that still samples \citep{tgi2026}, and provider APIs accept zero without saying which applies \citep{openai2026responses,deepseek2026api}.}

Deliberate sampling is the source most directly controlled by the researcher when the model provider exposes generation parameters. Setting temperature to zero can substantially improve consistency and has no direct monetary cost.

Several prominent reasoning models restrict this control in favor of a reasoning intensity, which lets users tune the effort and cost of a query while the provider manages sampling. As of November 2026, Anthropic tells users to omit temperature for Opus 5.5 \citep{anthropic2026opus55}, OpenAI tells them to remove it for GPT-6 models unless reasoning effort is none \citep{openai2026gpt6}, and Google states that Gemini 4 and Gemini 3.6+ ignore custom values \citep{google2026prompting}. DeepSeek also ignores temperature, without an error, unless the user disables its default thinking mode \citep{deepseek2026thinking}. In these cases the researcher cannot remove deliberate sampling directly, and a stated temperature may have had no effect. 


\subsection{Silent model updates}

A provider may alter the model behind an unchanged public name. The endpoint, prompt, and user parameters can remain the same while the underlying weights or safety guardrails change. For example, Anthropic redeployed Claude Fable 5 on July 1, 2026, with a new safety classifier that flags benign requests more often, and flagged requests are answered by Claude Opus 4.8 instead \citep{anthropic2026fable5}. Relatedly, \citet{chen2024} document substantial variation in the behavior of GPT-4 over time. In one simple mathematical task involving the classification of numbers as prime or composite, reported accuracy fell from 84 percent to 51 percent within three months. 

Software updates create familiar reproducibility problems, but model updates can be more severe. An older software version (e.g., Stata 15, released 2017) can often be archived and installed later. Though providers typically pin specific legacy models through their APIs, a proprietary model version may disappear completely. Even a dated model identifier may not reveal changes to system messages, filters, routing rules, or infrastructure.

Silent updates create both short-term and long-term risks. A model may change during data collection, so observations processed on different dates may not be comparable. It may also change after publication, preventing reproducibility verifiers from recovering the original outputs. Researchers using online services should therefore record the model identifier returned by every call, the date and time of execution, any provider fingerprint, and any available version information. These records do not ensure exact regeneration, but they make otherwise invisible changes easier to detect.

\subsection{Floating-point rounding}

Computers store numbers with finite precision, so they round the results of their calculations. Rounding makes the order of operations matter. In R, for example, \texttt{1 + (1e20 - 1e20)} returns 1, whereas \texttt{(1 + 1e20) - 1e20} returns 0, because the 1 is lost when it is added to the much larger number. An LLM performs a vast number of such operations, and their order or results can change between runs for three reasons:

\begin{enumerate}
\setlength\itemsep{-0.0em}
    \item \textit{Server load.} Online providers group requests into batches, which affects how operations are scheduled. The composition and size of the batch may differ across repeated calls.
    \item  \textit{Hardware.} Different processors, drivers, numerical libraries, and precision settings can implement the same operation differently. Providers may assign repeated requests to different machines and may also replace or upgrade this infrastructure without notifying users.
    \item \textit{Model splitting.} Large models are often distributed across several processors. Differences in scheduling and communication can alter the order in which partial results are combined.
\end{enumerate}

Numerical and parallel computing have long faced these problems. With LLMs, however, researchers often do not know and cannot control where the computations occur, which hardware performs them, or which other requests are processed at the same time.

Most changes in the final digits do not matter. If one token has a clearly higher logit than all others, small numerical differences will not alter the selected token. When the leading candidates are nearly tied, however, a tiny numerical change can reverse their order, and all later tokens are then generated under a different context.

Empirical evidence shows that outputs vary at zero temperature, but the causes remain debated. \citet{ouyang2024} evaluate 829 code generation tasks, find that repeated requests frequently produce different outputs at zero temperature, and list floating-point behavior, unreliable GPU calculations, and mixture-of-experts routing as competing hypotheses. \citet{he2025} identify batch non-invariance as the primary reason, showing that server-side batching can alter numerical computations even under zero temperature. 

\subsection{Expert routing}

Some LLMs use a mixture-of-experts architecture. Rather than using every part of the model for every token, the model sends each token to a small number of specialized components, known as experts. This allows the model to have greater overall capacity without a comparable increase in computing cost. However, each expert may be able to process only a limited number of tokens at a time. If too many tokens are sent to the same expert, some must either skip that expert or be sent elsewhere \citep{fedus2022switch}.

This creates a potential source of variation. Whether a token reaches its preferred expert can depend on which other users' requests happen to be processed in the same batch. As a result, two identical requests can, in principle, receive different outputs because they were processed alongside different requests. As part of a proof-of-concept adversarial attack, \citet{hayes2024} show that carefully chosen requests in the same batch can change a model's response to an otherwise unchanged input. The effect depends on how the model is implemented, which is generally not visible to the user.

\subsection{What researchers can control}

Table~\ref{tab:sources} sorts the sources of variation by whether researchers can act on them. Local execution of open weights provides the strongest available control. Researchers can archive the weights, fix the batch size, preserve the software environment, record the hardware, and request deterministic numerical operations where supported. Exact regeneration may then be achievable, although it requires control of the complete serving stack rather than the weights alone.

\begin{table}
\TABLE
{Differences between two runs of an identical prompt.\label{tab:sources}}
{\small
\begin{tabular}{@{}L{0.15\textwidth}L{0.53\textwidth}C{0.09\textwidth}C{0.08\textwidth}C{0.1\textwidth}@{}}
\toprule
\textbf{What differs between runs} & \textbf{In plain terms} & \textbf{Randomness survives $T{=}0$?} & \textbf{Control: local} & \textbf{Control: API} \\
\midrule
\multicolumn{5}{@{}l}{\textit{\textcolor{sciblue}{Deliberate randomness}}}\\
\addlinespace[3pt]
Sampling & The model draws each word at random from a probability distribution, instead of always taking the word with the highest logit. & \textbf{no} & full & for some models only \\
\addlinespace[6pt]
\multicolumn{5}{@{}l}{\textit{\textcolor{sciblue}{The model is not the same model}}}\\
\addlinespace[3pt]
Silent updates & The provider swaps the model behind an unchanged name. & yes & full & none \\
\addlinespace[6pt]
\multicolumn{5}{@{}l}{\textit{\textcolor{sciblue}{Rounding: same arithmetic, different order, different results}}}\\
\addlinespace[3pt]
Server load & The prompt is computed alongside whatever other prompts arrived at the same moment, which changes how the arithmetic is divided up. & yes & full & none \\
\addlinespace[2pt]
Hardware & A different chip, driver, or library version divides the same arithmetic differently. & yes & full & none \\
\addlinespace[2pt]
Model splitting & Very large models are cut across several chips, and the partial results are recombined in an order that varies. & yes & partial & none \\
\addlinespace[6pt]
\multicolumn{5}{@{}l}{\textit{\textcolor{sciblue}{A different part of the model ran}}}\\
\addlinespace[3pt]
Expert routing & Large models are a committee of experts, and each token is sent to a few of them. Places are limited, and tokens from other users' prompts can take them, so a token may go to its second choice. & yes & partial & none \\
\bottomrule
\end{tabular}}
{Note. \textit{Local} means the model runs on the user's hardware; \textit{API} means it runs on the provider's infrastructure.}
\end{table}

Commercial services provide much less control, especially when temperature cannot be set. Researchers generally have no control over the hardware, batching behavior, or silent model updates. Local execution of open-weight models avoids this loss of control. It eliminates the dependence on the continued availability and policies of a private provider \citep{spirling2023,ollion2024,palmer2024}. Open weights alone are not enough, however. \citet{higton2026} access eight open-weight models through a routing service with 33 hosting providers and find that only 26\% to 92\% of sentences receive the same label across 100 repetitions, with much of the variation within providers, whereas the same task returns identical labels at $T=0$ on their own hardware. Local execution nevertheless creates its own documentation requirements. The researcher must report the inference library, numerical precision, processor model, driver versions, batching settings, and any options related to deterministic execution.


\section{Illustration with annual corporate filings}\label{sec:illustration}

We illustrate the practical consequences of model variation through a common measurement task. We follow the lines of \cite{tetlock2008more} and \cite{loughran2011liability}, who regress stock returns on proxies for sentiment. Here, sentiment is derived from LLM judgment on 10-K filings. The model classifies the filings' tone as positive, neutral, or negative. The experiment has two purposes. First, it measures output variation in a setting representative of applied work. Second, it separates deliberate sampling from residual variation where the available model settings permit this distinction.

\subsection{Experimental design}

The source documents are excerpts from Item 1 and Item 1A of annual filings submitted by S\&P 500 firms for 2025 on the SEC's EDGAR platform. Each excerpt is stored as a plain text file identified by ticker symbol. To limit API costs, the sample is restricted to the 100 firms with the shortest excerpts by token count. This restriction is purely for computational convenience, as the objective is to illustrate the existence of output variability rather than to estimate a population quantity. The following prompt was used on July 20, 2026:

\vspace{3mm}

\noindent{\ttfamily\raggedright\fontsize{12}{12.5}\linespread{01}\selectfont
Below is an excerpt from a company's 10-K filing (Item 1 Business and Item 1A Risk Factors).
[EXCERPT]
Classify the overall tone of this passage from a company's 10-K filing as positive, negative, or neutral for the company. Answer with exactly one word: positive, negative, or neutral.\par}

The procedure is repeated 200 times with exactly the same prompt on the same model. This is carried out with three models: (1) OpenAI's GPT-5.6 Luna, (2) DeepSeek V4-Pro, and (3) Google's Gemma 4-26B-A4B. While DeepSeek has not publicly disclosed a training-data or knowledge cutoff, GPT-5.6 was trained on data through February 2026, and Gemma 4 on data through January 2025. Accordingly, our sample may have been included in the training data available to GPT-5.6, but not in that available to Gemma 4. This difference could contribute to variation in the models' outputs, although such analyses are beyond the scope of our paper.

For OpenAI, we set the reasoning effort to ``none" and do not set temperature. For DeepSeek, we disable thinking mode, because temperature otherwise has no effect. For models (2) and (3), we report results for both zero and nonzero temperatures. Sentiment scores are then digitized as follows: positive responses are coded as $1$, neutral responses as $0$, and negative responses as $-1$. For each repeated run, annual stock returns $r_{i,t}$ from the calendar year 2025 are regressed on the corresponding sentiment score $s_{i,t}$ (synchronous regression):
\begin{equation}
\label{eq:sentiment}
r_{i,t} = \alpha + \beta \times  s_{i,t} + e_{i,t}.
\end{equation}

The regression produces a coefficient $\hat{\beta}$, a standard error, and a $t$-statistic. It is the latter that we are interested in, because it captures both the sign and statistical significance of the coefficient.

\subsection{Results}

\begin{figure}
\FIGURE
{\includegraphics[width=0.8\textwidth]{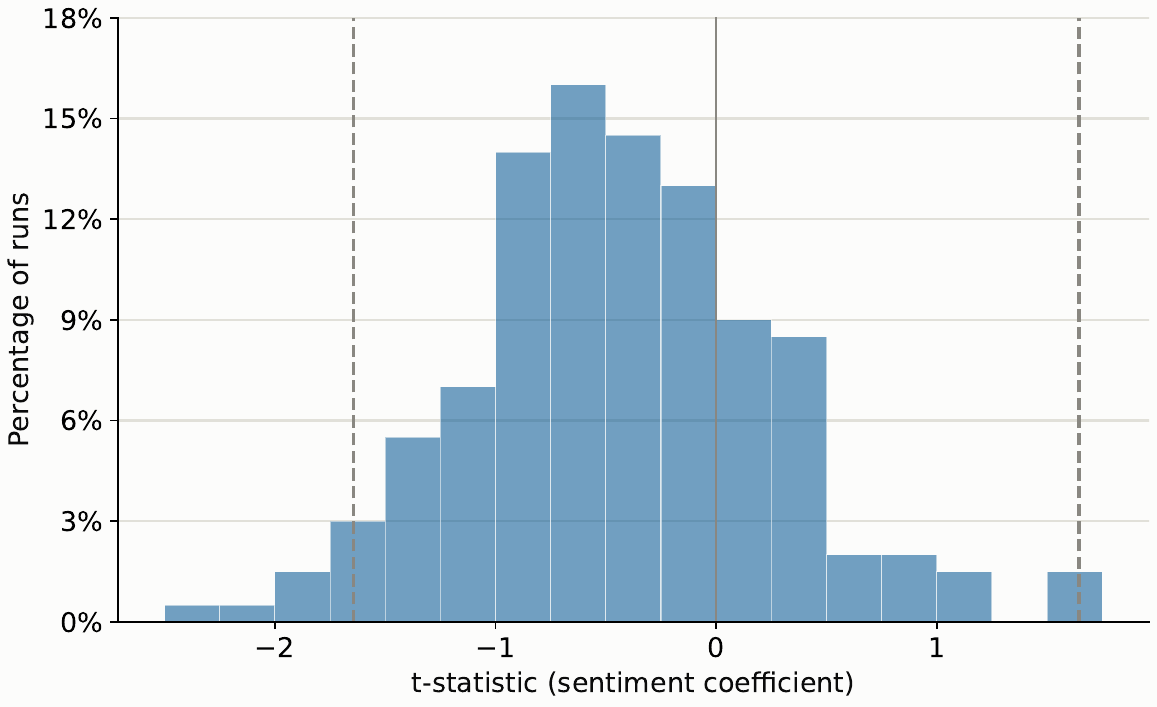}}
{Distribution of regression statistics across repeated OpenAI model draws.\label{fig:openai}}
{The figure displays the distribution of test statistics from the regressions defined in Equation \eqref{eq:sentiment} over the 200 runs. The two dashed vertical lines mark the critical values of a two-sided test at the 10\% level.}
\end{figure}

Figure \ref{fig:openai} shows the distribution of $t$-statistics associated with GPT-based sentiment scores. We leave temperature at the API default. OpenAI's API reference lists a default of 1 without saying whether it applies to reasoning models \citep{openai2026responses}. The generated scores vary across runs, even though the set of outcomes is small (only three possible values for each firm in our experiment). The model reads the content of the filings and interprets them in different ways for each successive run.

This first experiment shows that repeated classifications can produce very different regression statistics even though the filing excerpts, prompts, coding rules, and regression specification remain fixed. Around 5\% of the $t$-statistics fall below the lower critical value of a two-sided test at the 10\% level, and the sign of the coefficient also changes across runs. This dispersion shows the sensitivity of the estimate to annotation randomness. It does not establish false positives or an inflated Type I error: the true coefficient is unknown, and the repeated draws hold the firm sample fixed.

\begin{figure}
\FIGURE
{\begin{minipage}{0.48\textwidth}
\centering
\includegraphics[width=\textwidth]{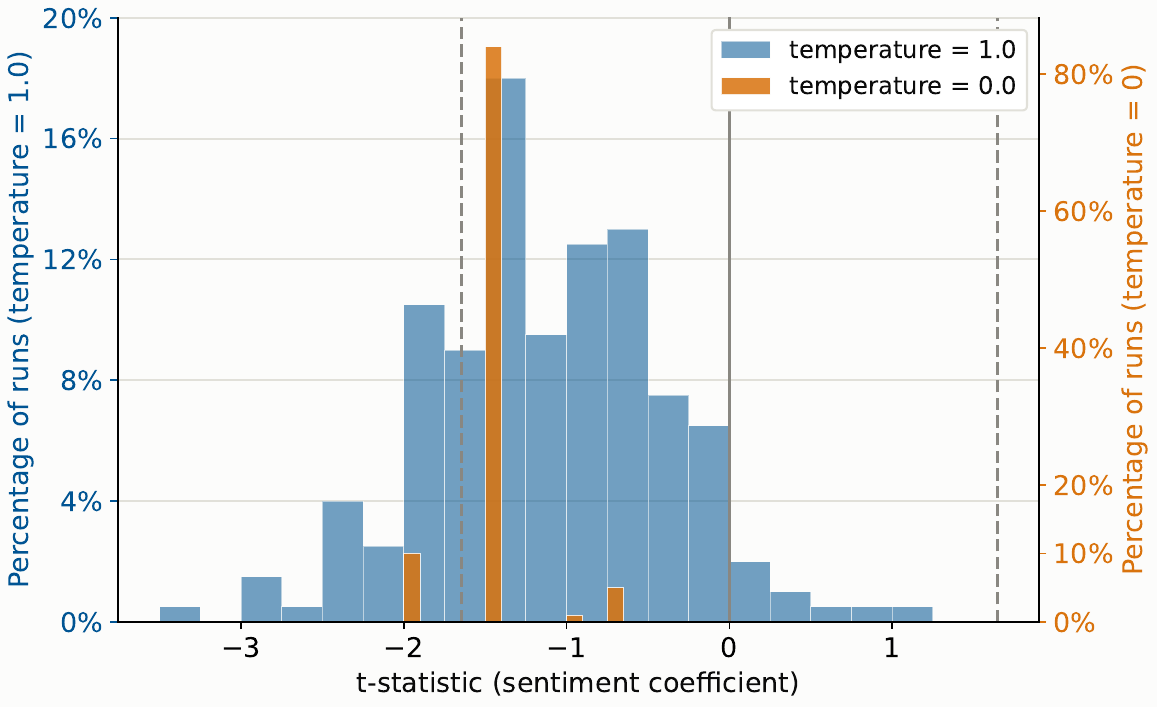}
\end{minipage}%
\begin{minipage}{0.505\textwidth}
\centering
\includegraphics[width=\textwidth]{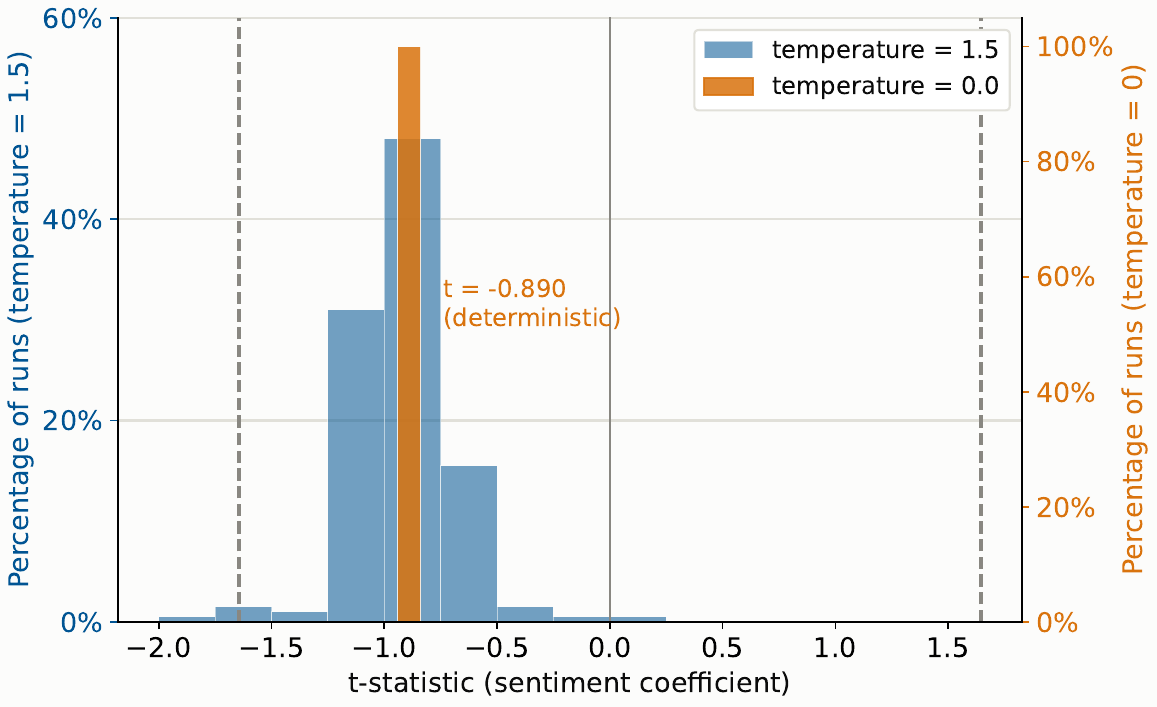}
\end{minipage}}
{Distribution of regression statistics across repeated DeepSeek and Gemma 4 model runs.\label{fig:deepseek}}
{The figure displays the distribution of test statistics from the regressions defined in Equation \eqref{eq:sentiment} over the 200 runs. The two dashed vertical lines mark the critical values of a two-sided test at the 10\% level. The bar color represents the chosen temperature. Light blue is used for $T>0$ (with frequencies reported on the left vertical axis) and green is used for $T=0$ (right vertical axis).}
\end{figure}

Figure \ref{fig:deepseek} repeats exactly the same exercise but in four different settings. First, we consider two alternative models: DeepSeek V4-Pro and Gemma 4-26B-A4B. The former is open-weight, but so large it can only be used via the provider's API. The latter is also open-weight and its smaller size makes it possible to run locally on a computer with at least 24 GB of RAM. Moreover, we carried out our tests with several temperature values: zero (to assess randomness under deterministic sampling) and the API default of one for DeepSeek (and 1.5 for Gemma).

The results exhibit markedly different degrees of dispersion. With unit temperature, the distribution of statistics based on DeepSeek-generated sentiment (left panel) is comparable to that obtained from OpenAI classifications in Figure \ref{fig:openai}. Under zero temperature, however, the distribution becomes much more concentrated but does not collapse to a single value. Instead, the estimates take only four distinct values (with one occurring only rarely). This residual variability arises because, among the 100 firms, two receive different sentiment classifications across repeated requests, yielding four possible combinations of outcomes. The result shows that setting $T=0$ removes deliberate sampling but, as explained in Section \ref{sec:randomness}, does not eliminate all sources of randomness in frontier models served through online APIs. By contrast, the distribution of statistics based on Gemma-4 deployed locally (right panel) is entirely concentrated at a single value under $T=0$. The residual variability disappears, indicating that this particular local configuration is internally deterministic.

This determinism is specific to the machine behind Figure \ref{fig:deepseek}. Repeating the full Gemma-4 protocol on eight further machines --- three Apple-Silicon chips and five NVIDIA GPU configurations --- shows that the $T{=}0$ point estimate itself is not stable across hardware: even the two most comparable machines in the set, running identical software and model weights, disagree on 2 of 100 firm labels.\footnote{Full results are reported in the appendix of the arXiv version of this paper (\url{https://arxiv.org/abs/2607.24372}). Across eight machines, the $T{=}0$ coefficient ranges from $-0.072$ to $-0.139$, and even the closest possible comparison --- two Apple-Silicon chips running identical software and model weights --- disagrees on 2 of 100 firm labels. At $T{=}1.5$, the across-run SD of the coefficient reaches up to 55\% of the within-run standard error on some machines (vs.\ under 25\% on others), up to 58 of 100 firms switch label at least once across 200 runs, and pairwise run agreement ranges from 97.6\% to 98.3\%.}

Our illustration has limits. It covers one task on 100 firms selected by excerpt length rather than at random, and one local configuration. Sentiment classification serves the purpose because it is familiar from finance and has three labels that make every disagreement between runs visible. Judgment-based tasks that need more context may vary more. Larger samples may dampen the effect of single disagreements on estimates, but they do not remove them. We did not test whether other batch sizes or numerical precisions yield the same Gemma labels, but Section~\ref{sec:randomness} gives reasons to expect that they may not; inference library is confounded with hardware in the test above, so we cannot isolate its effect either. We did not validate the labels against human judgments.

\section{Practical guidance for reproducible science}\label{sec:guidance}

Economics and management have made substantial progress in research transparency during the past two decades. Data availability policies, registered reports, replication archives, journal reproducibility verification, and third-party certification have improved the credibility of published evidence \citep{christensen2018,perignon2019,vilhuber2020,perignon2024a}. \textit{Management Science} introduced a Code and Data Disclosure policy in 2019 \citep{fivsar2024reproducibility} and has since strengthened its reproducibility requirements through a partnership with the \href{https://www.cascad.tech}{Cascad} certification agency to independently verify the computational reproducibility of accepted articles \citep{loch2025reinforcing}. In April 2026, the journal introduced an additional requirement for authors to disclose on AsCollected.org when and how their data were collected \citep{mnsc2026policy}. More broadly, the American Economic Association policy, the Social Science Data Editors template README, and the Data and Code Availability Standard (DCAS), currently endorsed by 19 journals, provide widely used standards for replication packages \citep{vilhuber2019,vilhuber2022,koren2022}.

These policies and templates require data, code, access instructions, provenance, and information about the computational environment. They do not yet provide sufficiently explicit guidance for model-generated variables. The closest standard is GUIDE-LLM, a consensus checklist for behavioral and social science \citep{feuerriegel2026}. Its 14 core items cover the model version, access route, configuration, number of runs, prompts, human validation, and code. They do not ask authors to deposit raw outputs or report how much results vary across runs, and they do not say what journals should verify. These gaps matter because conventional empirical pipelines already reproduce imperfectly at scale \citep{perignon2024b}, and generative models add another layer of complexity. Our guidance applies to any model whose output enters the data and can differ across runs or over time. Tools with stable output, such as dictionaries or a versioned classifier run locally, need only conventional version reporting. The provider details in Section~\ref{sec:randomness} will age, but the items in Table~\ref{tab:reporting} do not depend on them.

Of course, reproducibility in itself tells us nothing about validity. Gemma at $T=0$ returns the same labels in every run of Figure~\ref{fig:deepseek}, but this says nothing about whether the labels capture tone. Human coding, in turn, can be valid yet noisy, and researchers handle this noise by reporting agreement among independent coders. Repeated runs only measure a model's agreement with itself. Depending on the application, agreement with human judgments or across different models can also be informative, and whether such a check is needed is best judged in the review process.

\subsection{What authors need to report}\label{sec:what-to-report}

Authors should report details in two places. First, the paper should give readers what they need to assess the results: which model produced the data, under which settings, and how stable the output is. Second, the replication package should provide what other researchers or reproducibility verifiers need to reproduce them: the exact machine-readable artifacts, above all every model input and raw output before any post processing. Prompts belong here too, because prompts are code; a screenshot is not an adequate substitute, since it cannot be executed, searched, or compared. Where variation is the goal, as in simulated survey respondents, a nonzero temperature may be needed. In such cases, authors should report both the temperature and the number of draws, depositing all draws in the replication package.

Table~\ref{tab:reporting} classifies each relevant item along both dimensions and by priority. Items marked as \textit{must haves} are essentially free to provide, and no paper using model-generated data should omit them. The exception is the variability report, which requires repeated runs and is mandatory only when LLM-generated measurements are central to the findings. A measurement is central if it serves as the outcome, a main explanatory variable, or a sample filter in a result on which the paper's conclusions rest. The \textit{good to have} items involve archiving or validation; they entail a cost, which the experiment in Section~\ref{sec:illustration} makes concrete.

\begin{table}
\TABLE
{Reporting items for model generated variables.\label{tab:reporting}}
{\small
\begin{tabular}{@{}L{0.44\textwidth}L{0.27\textwidth}L{0.27\textwidth}@{}}
\toprule
\textbf{Item} & \textbf{In the paper} & \textbf{In the replication package} \\
\midrule
\multicolumn{3}{@{}l}{\textit{\textcolor{sciblue}{What the model was}}}\\
\addlinespace[3pt]
Model identification: name, exact version, provider, open or proprietary, access route (including any routing service and host), date when the model was queried & Must have & Must have (including hardware and serving stack for local execution) \\
\addlinespace[4pt]
Archived open weights and their cryptographic hash & n/a & Good to have \\
\addlinespace[2pt]
Training data cutoff and its relation to the sample period & Must have (if available) & n/a \\
\addlinespace[6pt]
\multicolumn{3}{@{}l}{\textit{\textcolor{sciblue}{How it was run}}}\\
\addlinespace[3pt]
Verbatim prompts, system messages, and examples & Must have (main prompt in an appendix) & Must have (prompts are code) \\
\addlinespace[4pt]
Generation parameters, distinguishing explicitly set values from provider defaults & Must have (headline settings, especially temperature) & Must have (complete configuration) \\
\addlinespace[6pt]
\multicolumn{3}{@{}l}{\textit{\textcolor{sciblue}{What it produced}}}\\
\addlinespace[3pt]
All model inputs and raw outputs from every run, before processing & n/a & Must have\\
\addlinespace[4pt]
Variability report, number of draws, agreement measures, sensitivity of main estimates & Must have if central, else good to have (distribution of results for central claims) & Must have if central (code that computes the report from the deposited draws) \\
\addlinespace[4pt]
Human validation of outputs used as measurements & Good to have & Good to have (human labels) \\
\addlinespace[6pt]
\multicolumn{3}{@{}l}{\textit{\textcolor{sciblue}{How much it cost}}}\\
\addlinespace[3pt]
Cost statement: estimated token counts, monetary cost, runtime & n/a & Must have \\
\bottomrule
\end{tabular}} 
{
}
\end{table}

Much of Table~\ref{tab:reporting} already fits the current \textit{Management Science} policy and its replication package checklist \citep{mnsc2024checklist, mnsc2026policy}. Prompts and generation parameters are part of the code that produces the raw data, which the checklist requires. Model identifiers and query dates are software and data versions, archived weights resemble licensed code, and the cost statement extends the run time the checklist already asks for. Human validation, like the training data cutoff, is for reviewers to assess rather than for the checklist. Three extensions would close most of the remaining gap at little burden: record each model as a data source in the provenance disclosure, make raw model outputs compulsory rather than encouraged because they cannot be regenerated, and require the variability report where LLM-generated measurements are central.

No other INFORMS journal addresses model-generated data, and most require less than \textit{Management Science}. \textit{Operations Research} also expects code and data, at the time of submission, but checks them only in voluntary replication studies \citep{opre2026policy, opre2026instructions}. \textit{Marketing Science} requires them after acceptance, but accepts a random subset or synthetic data in place of the original data \citep{desai2013}. \textit{Manufacturing \& Service Operations Management} and \textit{Information Systems Research} require no replication package, only that authors retain raw data and provide them on request \citep{msom2026ethics, isre2026ethics}. The same three extensions would apply to these journals.


Cost is a legitimate concern, especially for large input data, but it does not justify ignoring variation. Repeated calls on a documented random subsample can provide an affordable assessment. Authors should explain the choice of repetition strategy and sample size in relation to the observed instability and the importance of the affected result.

\subsection[What editors, reviewers, and data editors check]{What editors, reviewers, and data editors need to check}\label{sec:how-to-verify}

The replication package column of Table~\ref{tab:reporting} supports three assessments of an LLM-based pipeline. \textit{Reproduction} is an assessment of whether the deposited outputs yield the published results. This is typically checked by data editors and it is usually a preliminary step for users of replication packages. \textit{Replication} evaluates whether outputs newly generated for the same sample yield estimates consistent with the original distribution. \textit{Robustness} assesses whether the substantive conclusions survive variability in model outputs. The latter two should be evaluated by editors and referees, especially when central.

Reproduction uses the archived raw outputs. A data editor or reproducibility reviewer skips model execution and runs the remaining parsing, cleaning, and statistical analysis. This check should reproduce the published results exactly, subject to conventional software and numerical considerations. If it fails, the cause lies in the package and the authors must fix it. The raw output deposit is the most important item in the package because this check works even after the model has changed, becomes too expensive to run, or disappears altogether.

Replication rests on the deposited prompts, generation parameters, and execution instructions. The data generation itself is repeated. With a commercial service, the new output may differ from the archive, so it should be read as a fresh model draw rather than an exact copy of the archived output. The variability report supplies the acceptance criterion: an estimate within the documented distribution is consistent with the original, while one outside it signals a possible model update, infrastructure change, or underestimated variation. Such a failure does not by itself invalidate the published result, because the model may have changed. The README file should give an expected runtime and cost for regeneration, which verification teams and certification services need to plan and price their services. Replication usually lies beyond the checks conducted by data editors or certification agencies. However, editors and referees may require variability reports when central. In those cases, the raw outputs from each model generation should also be deposited.

Robustness uses the repeated draws in the deposited outputs and needs no new model runs. Authors assess whether the main estimates support the same conclusion across draws and report the result. Editors and reviewers judge whether the conclusions hold, and data editors verify that the reported figures follow from the deposited draws. If conclusions change across draws, the paper should say so. Data editors do not engage with robustness (or replication) beyond what is reported in the paper and deposited in the package.

Replication and robustness can disagree. In Figure \ref{fig:openai}, a new draw would likely fall within the documented distribution, so replication would succeed, yet the sign and significance of the coefficient change across draws, so the conclusion is not robust. Conversely, a model update can move estimates outside a narrow distribution, so replication fails even though the original conclusion is robust to output variability.

\section{Conclusion}\label{sec:Conclusion}


The recent arrival of LLMs has opened many promising avenues for research, including new ways of generating data. Yet this growing use of LLMs raises a fundamental reproducibility challenge because their outputs are inherently non-deterministic. As the adoption of LLMs for data generation has accelerated, research practices for understanding, documenting, and addressing these sources of randomness have not kept pace. This paper is our attempt to start a serious conversation about remedying this situation.

We document deliberate sampling, silent updates, server batching, hardware differences, model splitting, and expert routing as sources of non-determinism in LLM outputs. Some of these sources can be controlled by the user, while others remain outside the user's control, such as the deployment and order of computations on remote computing infrastructure with unknown and continuously evolving technical specifications. The use of open-weight models has important advantages over proprietary models, not least because commercial model providers do not commit to preserving legacy models for future use. Nevertheless, even open-weight models present reproducibility challenges and require great care in documenting the computational environment in which the model was deployed. Taken together, these considerations lead us to view LLM outputs as draws from a distribution rather than as fixed measurements.

The main contribution of this paper is practical guidance on what authors should report about the variability of LLM-generated data, what editors and reviewers should assess, and what data editors should verify. When outputs serve as measurements, researchers should set temperature to zero where possible, while avoiding the claim that this makes output deterministic. Tasks that seek variation, such as simulated respondents, may need a nonzero temperature. Researchers should also recognize that exact regeneration depends critically on where and how the model is run: commercial services generally cannot guarantee it, whereas archived open weights on a controlled local stack offer the strongest prospect of achieving it if the full serving environment is documented and preserved, potentially through containerization. Journals, in turn, should require documentation of the model and version, tuning parameters, machine-readable prompts, query dates, and all generated outputs as raw data in the replication package. Replication packages should support both reproduction from the archived outputs and replication from the original environment and prompts, and authors should report whether their conclusions are robust to variability in model outputs. Where LLM-generated measurements are central to the findings, a variability report should be required. Most of these requirements fit existing policies: prompts are code, model outputs are raw data, and the provenance disclosure can record each model as a data source. 



We share the profession's excitement about the promise of LLMs for research and expect that many important findings will come from their judicious use. Researchers often state that \textit{standing on the shoulders of giants} is what enabled their particular advance. We want to caution that certain steps must be taken urgently to ensure that the foundations provided by this new class of giants are stable enough to build upon.

\clearpage

\begin{APPENDICES}
\ifarxiv

\section{Hardware Experiment}\label{sec:hardware}

The Gemma results in Section \ref{sec:illustration} use pinned, open model weights executed locally, which removes provider-side updates and routing as sources of variation. This leaves one remaining channel that a proprietary API cannot isolate: the hardware and numerical backend performing the computation. We re-ran the full Gemma-4 protocol (same 100 firms, same filings, same prompts, same returns data, $T=0$ and $T=1.5$, 200 runs per firm per temperature) on eight machines spanning three Apple Silicon chips (M1 Pro, M4, M4 Max, all MLX backend) and four NVIDIA GPU configurations (A40, A40 $\times$2 with the model split across cards, RTX 3060 $\times$2, RTX 6000 Ada 49GB, all CUDA backend with \texttt{bitsandbytes} 4-bit quantization). The software environment is pinned to identical package versions across machines, documented in the replication package README and archived at \url{https://doi.org/10.5281/zenodo.24198290}; the only difference across machines is the chip and the corresponding backend. MLX is Apple's array framework for Apple Silicon's unified memory architecture; CUDA is NVIDIA's general-purpose GPU computing platform, here combined with \texttt{bitsandbytes} for 4-bit quantized inference. The two backends implement the same mathematical operations with different kernel code, rounding order, and quantization routines.

Our metrics of interest are the ones from Section \ref{sec:illustration}: the sentiment-tone coefficient estimated at $T=0$ and the mean and standard deviation of that coefficient across 200 repeated runs at $T=1.5$. At $T=0$ the coefficient should be identical across runs on a given machine (Gemma's $T=0$ decoding is internally deterministic); any spread would indicate a failure of that determinism, while any disagreement \textit{across} machines isolates the effect of hardware alone.

\begin{table}[h]
\TABLE
{Gemma-4 results on different hardware.\label{tab:hardware}}
{\small
\begin{tabular}{llcc}
\toprule
Hardware & Backend & $T{=}0$ coefficient (SD) & $T{=}1.5$ mean coefficient (SD) \\
\midrule
PTL11077 (M1 Pro) --- reference & MLX & $-0.1224$ (0) & $-0.1245$ (0.0261) \\
PCPEM52119 (M4) & MLX & $-0.1386$ (0) & $-0.1335$ (0.0351) \\
il-lv39 (M4 Max) & MLX & $-0.1386$ (0) & $-0.1361$ (0.0315) \\
cbsugpu06 (A40) & CUDA & $-0.0718$ (0) & $-0.104$ (0.113) \\
cv-ai (RTX 3060 $\times$2) & CUDA & \multicolumn{1}{c}{$^{a}$} & $-0.0505$ (0.129) \\
cbsugpu06 (A40 $\times$2, split) & CUDA & $-0.0718$ (0)$^{b}$ & $-0.104$ (0.113) \\
cbsugpu10 (RTX 6000 Ada) & CUDA & $-0.0718$ (0) & $-0.088$ (0.100) \\
cv-ai (RTX 3060 $\times$2, re-run) & CUDA & \multicolumn{1}{c}{$^{a}$} & $-0.051$ (0.129) \\
\bottomrule
\end{tabular}}
{\footnotesize $^{a}$No estimate: all 20,000 $T{=}0$ outputs on this card are classified \textit{neutral}, so tone has zero variance and the regression is not identified. Reproduced byte-identically on a re-run, so this is a stable property of the card/driver, and not a fluke. $^{b}$Byte-identical to the single-A40 run; splitting the model across two cards is not the cause of the RTX 3060's zero variance.}
\end{table}

Two findings stand out. First, at $T=0$ every machine returns a point estimate with zero within-machine spread (where identified), confirming that Gemma's deterministic decoding path is deterministic \textit{on a given machine}; the RTX 3060 is the exception, and it fails for a reason unrelated to randomness: its $T=0$ outputs collapse to a single label for every firm. Second, the $T=0$ point estimate itself is not stable \textit{across} machines: it ranges from $-0.0718$ (all CUDA cards) to $-0.1386$ (Apple M4 and M4 Max), with the reference M1 Pro in between at $-0.1224$. Backend (MLX versus CUDA/\texttt{bitsandbytes}) explains most of this split, but not all of it: the three MLX machines do not agree with each other either, and the closest possible comparison in the campaign -- the M1 Pro reference against the M4, same backend, same quantized weight revision, same \texttt{mlx} and \texttt{transformers} versions, same input data -- still disagrees on 2 of the 100 firms (NSC and VRSN), each flipping to the opposite label on the other machine, consistently across all 200 runs on each side. With software held fixed, two firms land on opposite sides of a near-tied next-token decision purely as a function of the chip. Before attributing the NSC/VRSN disagreement to hardware, we ruled out the mundane explanations: a missing row, a duplicated or mismatched input file, or a stale Yahoo returns pull. Table \ref{tab:audit} summarizes this check.

\begin{table}[h]
\TABLE
{Audit of the reference (M1 Pro) vs.\ M4 disagreement.\label{tab:audit}}
{\small
\begin{tabular}{@{}L{0.32\textwidth}L{0.58\textwidth}@{}}
\toprule
Check & Result \\
\midrule
Input filings (NSC.txt, VRSN.txt) & Identical, single shared file --- not duplicated per run \\
Yahoo returns data & Identical, single shared file --- not duplicated per run \\
Model weights/revision & Identical (\texttt{6b06e97d84a3}) \\
Firm count, $T{=}0$ row count & Identical: 100 firms, 20{,}000 rows each \\
Internal consistency (200 runs/firm at $T{=}0$) & Perfect: each firm's output is 100\% stable across all 200 runs on a given machine \\
Actual divergence & Exactly 2 firms, NSC and VRSN --- deterministic 100\%-consistent flip in opposite directions on each machine \\
\bottomrule
\end{tabular}}
{}
\end{table}

Given that we can rule out data-pipeline errors, like missing data or other mismatched inputs, we conclude that with the software stack held essentially fixed, two firms still land on opposite sides of a near-tied next-token logit purely as a function of the chip they ran on.

We then extended the same two diagnostics used for the main-text models in Section \ref{sec:illustration} --- the ratio of across-run SD of the coefficient to the mean within-run SE, and the label-instability counts $K$ (firms ever taking more than one label across the 200 runs), $k$ (mean per-run disagreement with each firm's own majority-vote label), and pairwise run agreement --- to all eight machines in the hardware campaign. Table \ref{tab:hardware-variance} reports the SD/SE ratio, and Table \ref{tab:hardware-label} reports label instability, both split by temperature.

\begin{table}[h]
\TABLE
{SD(coef) across runs relative to mean within-run SE, by machine.\label{tab:hardware-variance}}
{\small
\begin{tabular}{lcc}
\toprule
Machine & SD(coef)/mean SE, $T{=}0$ & SD(coef)/mean SE, $T{=}1.5$ \\
\midrule
PTL11077 (M1 Pro) --- reference & $<0.001$ & 0.183 \\
PCPEM52119 (M4) & $<0.001$ & 0.247 \\
il-lv39 (M4 Max) & $<0.001$ & 0.222 \\
cbsugpu06 (A40) & $<0.001$ & 0.486 \\
cv-ai (RTX 3060 $\times$2) & ---$^{a}$ & 0.550 \\
cbsugpu06 (A40 $\times$2) & $<0.001$ & 0.486 \\
cbsugpu10 (RTX 6000 Ada) & $<0.001$ & 0.424 \\
cv-ai (RTX 3060 $\times$2, re-run) & ---$^{a}$ & 0.550 \\
\bottomrule
\end{tabular}}
{\footnotesize $^{a}$No tone variance at $T{=}0$ on this card (all outputs \textit{neutral}), so the ratio is undefined.}
\end{table}

\begin{table}[h]
\TABLE
{Label instability by machine: $K$ = firms ever taking more than one label across 200 runs (/100); $k$ = mean per-run disagreement with each firm's own majority-vote label; agreement = mean pairwise run agreement.\label{tab:hardware-label}}
{\small
\begin{tabular}{lcccccc}
\toprule
Machine & $K$ ($T{=}0$) & $k$ ($T{=}0$) & Agreement \% ($T{=}0$) & $K$ ($T{=}1.5$) & $k$ ($T{=}1.5$) & Agreement \% ($T{=}1.5$) \\
\midrule
PTL11077 (M1 Pro) --- reference & 0 & 0.0 & 100.0 & 25 & 1.69 & 97.8 \\
PCPEM52119 (M4) & 0 & 0.0 & 100.0 & 31 & 0.99 & 98.3 \\
il-lv39 (M4 Max) & 0 & 0.0 & 100.0 & 31 & 0.96 & 98.3 \\
cbsugpu06 (A40) & 0 & 0.0 & 100.0 & 54 & 1.31 & 97.8 \\
cv-ai (RTX 3060 $\times$2) & 0 & 0.0 & 100.0 & 52 & 1.48 & 97.6 \\
cbsugpu06 (A40 $\times$2) & 0 & 0.0 & 100.0 & 54 & 1.31 & 97.8 \\
cbsugpu10 (RTX 6000 Ada) & 0 & 0.0 & 100.0 & 58 & 1.47 & 97.7 \\
cv-ai (RTX 3060 $\times$2, re-run) & 0 & 0.0 & 100.0 & 52 & 1.48 & 97.7 \\
\bottomrule
\end{tabular}}
{}
\end{table}

At $T=0$ every machine is internally perfectly stable ($K=0$, 100\% agreement on every card, including the RTX 3060 once its degenerate all-\textit{neutral} output is taken as its own stable label). The informative column is $T=1.5$: label churn scales with the SD/SE ratio across machines, and both rise together moving from the three MLX (Apple Silicon) machines to the four CUDA cards. The two RTX 3060 runs, taken three days apart (PR \#47), are essentially identical to each other on every diagnostic, which rules out a one-off fluke and confirms that each card/driver combination produces a stable, reproducible --- but hardware-specific --- degree of output variability. Backend (MLX vs.\ CUDA/\texttt{bitsandbytes}) is therefore associated with how much a rerun moves the result, not only with the point estimate: the same software stack run on different chips does not just occasionally disagree, it disagrees at a rate that is itself a persistent property of the hardware.

Table \ref{tab:r1comment8} collects the same diagnostics across models, comparing a deterministic ($T=0$) and a stochastic ($T>0$) setting for each. The first pair of columns reports the standard deviation of the estimated coefficient across repeated runs relative to the mean within-run standard error, the second pair the number of firms (out of 100) that ever take more than one label, and the third pair the mean pairwise agreement between runs. At $T>0$ run-to-run noise is of the same order as sampling uncertainty (SD/SE of 0.69 for DeepSeek-v4-pro and GPT-5.6-luna, 0.29 for Gemma-4-26b), and between a quarter and 70\% of firms flip label at least once. Setting $T=0$ removes nearly all of this churn --- Gemma-4-26b becomes internally deterministic and DeepSeek-v4-pro reaches 99.7\% pairwise agreement --- but it does not eliminate it: DeepSeek's 200 nominally identical $T=0$ runs still produce four distinct (coefficient, $t$-statistic) pairs, spanning 0.19 in the coefficient and 1.24 in the $t$-statistic, and all of that spread traces back to just 2 of the 100 firms whose label flips. Even a fully pinned-down temperature therefore leaves a residual channel through which the same prompt and the same data can deliver a different estimate.

\begin{table}[h]
\centering
\caption{Variance and label-instability diagnostics across repeated runs, by model.}
\label{tab:r1comment8}
\small
\begin{tabular}{lcccccc}
\toprule
& \multicolumn{2}{c}{SD(coef) / mean within-run SE} & \multicolumn{2}{c}{Firms ever flip label (/100)} & \multicolumn{2}{c}{Pairwise run agreement (\%)} \\
\cmidrule(lr){2-3} \cmidrule(lr){4-5} \cmidrule(lr){6-7}
Model & $T{=}0$ & $T{>}0$ & $T{=}0$ & $T{>}0$ & $T{=}0$ & $T{>}0$ \\
\midrule
DeepSeek-v4-pro$^{a}$ & 0.26 & 0.69 & 2 & 69 & 99.7 & 95.4 \\
GPT-5.6-luna$^{b}$ & -- & 0.69 & -- & 71 & -- & 82.8 \\
Gemma-4-26b$^{c}$ & $<0.001$ & 0.29 & 0 & 25 & 100.0 & 97.8 \\
\bottomrule
\end{tabular}
\par\smallskip
\footnotesize
$^a$DeepSeek's $T{>}0$ column is the API default (temperature omitted). $\quad$
$^b$GPT-5.6 has no $T{=}0$ leg (no explicit temperature control); its one condition is shown under $T{>}0$. $\quad$
$^c$Gemma's $T{>}0$ column is $T{=}1.5$; its $T{=}0$ leg is internally deterministic.

\smallskip
\noindent DeepSeek's 200 ``identical'' $T=0$ runs collapse to 4 distinct (coefficient, $t$-statistic) pairs: $(-0.279,\,-1.79)$ in 20 runs, $(-0.192,\,-1.42)$ in 168 runs, $(-0.158,\,-0.82)$ in 2 runs, $(-0.085,\,-0.55)$ in 10 runs -- a range of 0.19 in the coefficient and 1.24 in the $t$-statistic, traced to the 2 firms (of 100) whose label flips at $T=0$.
\end{table}

\section{List of models}\label{sec:model_list}

The 212 models considered in Section \ref{sec:researcher_use} are listed below, grouped by providers.

\begin{multicols}{2}
\begin{itemize}
 \small \setlength\itemsep{-0.4em}
  \item \textbf{01.AI}: Yi-34B, Yi-1.5, Yi-Large.
  \item \textbf{Academic}: BioBERT, FinBERT, ClinicalBERT, Legal-BERT.
  \item \textbf{AI21}: Jurassic-2, Jamba.
  \item \textbf{Alibaba}: Qwen, Qwen-7B, Qwen-VL, Qwen1.5, Qwen2, Qwen2.5, QwQ, Qwen2.5-Max, Qwen3, Qwen3.5, Tongyi Qianwen.
  \item \textbf{Allen Institute for AI}: SciBERT.
  \item \textbf{Amazon}: Amazon Titan, Amazon Nova.
  \item \textbf{Anthropic}: Claude 1, Claude 2, Claude Instant, Claude 3, Claude 3 Opus, Claude 3 Sonnet, Claude 3 Haiku, Claude 3.5 Sonnet, Claude 3.5 Haiku, Claude 3.7 Sonnet, Claude Opus 4, Claude Sonnet 4, Claude Opus 4.1, Claude Sonnet 4.5, Claude Haiku 4.5, Claude Opus 4.5, Claude Opus 4.6, Claude Opus 4.7, Claude Opus 4.8, Claude Sonnet 4.6, Claude Sonnet 5, Claude Fable 5.
  \item \textbf{Baichuan}: Baichuan, Baichuan2.
  \item \textbf{Baidu}: ERNIE 3.0, ERNIE Bot, ERNIE 4.0, ERNIE 4.5, ERNIE 5.0, ERNIE 5.1, Wenxin Yiyan.
  \item \textbf{BigCode}: StarCoder.
  \item \textbf{BigScience}: BLOOM language model.
  \item \textbf{ByteDance}: Doubao, Doubao 2.0, ByteDance Skylark.
  \item \textbf{Cohere}: Cohere Command, Cohere Command R, Cohere Command R+, Cohere Command A, Cohere Aya.
  \item \textbf{Databricks}: DBRX.
  \item \textbf{DeepSeek}: DeepSeek LLM, DeepSeek-Coder, DeepSeek-V2, DeepSeek-V3, DeepSeek-VL, DeepSeek-Math, DeepSeek-R1, DeepSeek-V3.1, DeepSeek-V3.2, DeepSeek-V4.
  \item \textbf{EleutherAI}: GPT-NeoX, GPT-J, Pythia.
  \item \textbf{Google}: Gopher, Chinchilla, LaMDA, Flan-T5, PaLM 2, Med-PaLM, Med-PaLM 2, Google Bard, Gemini 1.0, Gemini 1.5, Gemini 2.0, Gemini 2.5, Gemini 3, Gemini 3.5, Gemini 3.6, Gemma, Gemma 2, Gemma 3, Gemma 4, CodeGemma, BERT, ALBERT, ELECTRA, XLNet, mBERT.
  \item \textbf{Huawei}: Huawei PanGu.
  \item \textbf{Hugging Face}: DistilBERT.
  \item \textbf{iFlytek}: iFlytek Spark, SparkDesk.
  \item \textbf{Inspur}: Yuan 2.0.
  \item \textbf{Kunlun}: Skywork.
  \item \textbf{LMSYS/Academic}: Vicuna.
  \item \textbf{Meta}: OPT-175B, Galactica, LLaMA, Llama 2, Code Llama, Llama 3, Llama 3.1, Llama 3.2, Llama 3.3, Llama 4, Muse Spark, RoBERTa, XLM-R, XLM-RoBERTa.
  \item \textbf{Microsoft}: Phi-1, Phi-2, Phi-3, Phi-3.5, Phi-4, WizardLM, Orca, DeBERTa.
  \item \textbf{MiniMax}: MiniMax-Text, MiniMax-01, MiniMax M3.
  \item \textbf{Mistral}: Mistral 7B, Mixtral 8x7B, Mixtral 8x22B, Mistral Large, Mistral Large 2, Mistral Small, Mistral Medium, Mistral Nemo, Codestral, Mathstral, Pixtral, Ministral, Mistral Small 3, Mistral Medium 3, Magistral, Mistral Large 3, Mistral Small 4, Mistral Medium 3.5, Ministral 3.
  \item \textbf{Moonshot AI}: Kimi Moonshot, Kimi K1.5, Kimi K2, Kimi K2.5.
  \item \textbf{MosaicML}: MPT-7B.
  \item \textbf{NVIDIA}: Nemotron, Megatron-LM.
  \item \textbf{OpenAI}: GPT-2, GPT-3, InstructGPT, GPT-3.5, ChatGPT, GPT-4, GPT-4 Turbo, GPT-4V, GPT-4o, GPT-4o mini, GPT-4.1, GPT-4.5, OpenAI o1, o1-mini, OpenAI o3, o3-mini, o4-mini, GPT-5, GPT-5.1, GPT-5.2, GPT-5.4, GPT-5.5, GPT-5.6.
  \item \textbf{SenseTime}: SenseChat, SenseNova.
  \item \textbf{Shanghai AI Lab}: InternLM, InternLM2, InternLM2.5.
  \item \textbf{Stability AI}: StableLM.
  \item \textbf{Stanford/Academic}: Alpaca language model.
  \item \textbf{StepFun}: StepFun.
  \item \textbf{Tencent}: Hunyuan, Hunyuan-Large, Hunyuan-T1.
  \item \textbf{TII}: Falcon-40B, Falcon-180B.
  \item \textbf{xAI}: Grok-1, Grok-1.5, Grok 2, Grok 3, Grok 4, Grok 4.5.
  \item \textbf{Zhipu AI}: GLM-130B, ChatGLM, ChatGLM2, ChatGLM3, GLM-4, GLM-4.5, GLM-5, GLM-5.2, CogVLM.
\end{itemize}
\end{multicols}
\fi

\clearpage
\ifarxiv
    
\section{Open versus closed models}\label{sec:open_closed}

\begin{figure}[!h]
\FIGURE
{\includegraphics[width=0.9\textwidth]{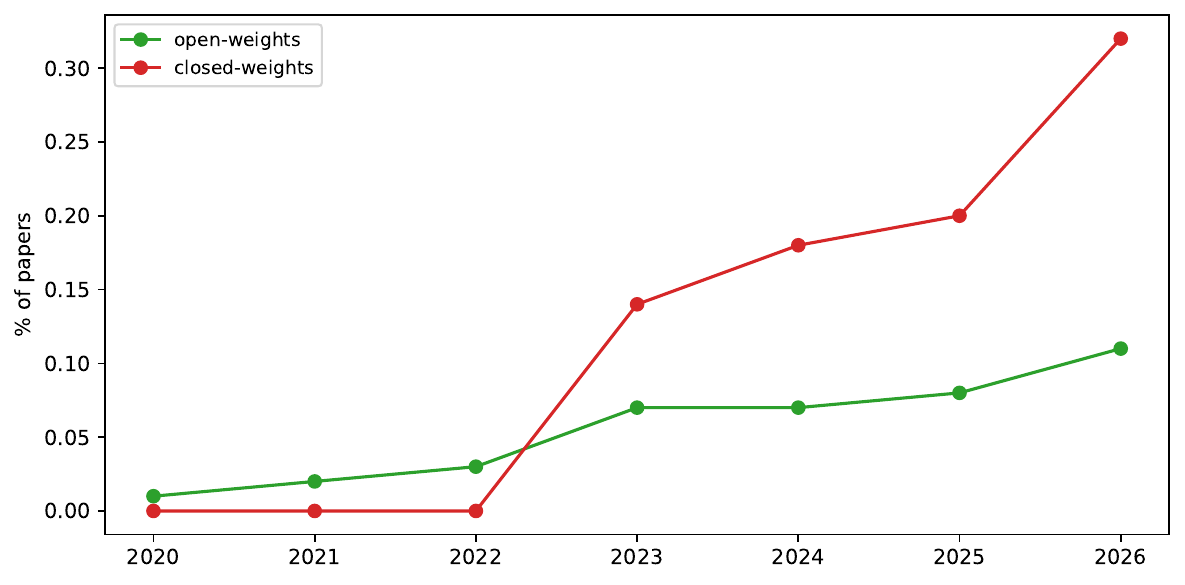}}
{Open versus closed LLM models.\label{fig:open_closed}}
{We chronologically represent the proportion of models that are mentioned in titles and abstracts, grouped according to whether they are open or closed. The total number of articles published in the 145 journals are 15,545 (2020), 15,615 (2021), 14,573 (2022), 13,700 (2023), 14,185 (2024), 14,664 (2025), and 13,610 (2026).}
\end{figure}

\section{LLM provider market shares}\label{sec:provider_share}

\begin{figure}[!h]
\FIGURE
{\includegraphics[width=0.9\textwidth]{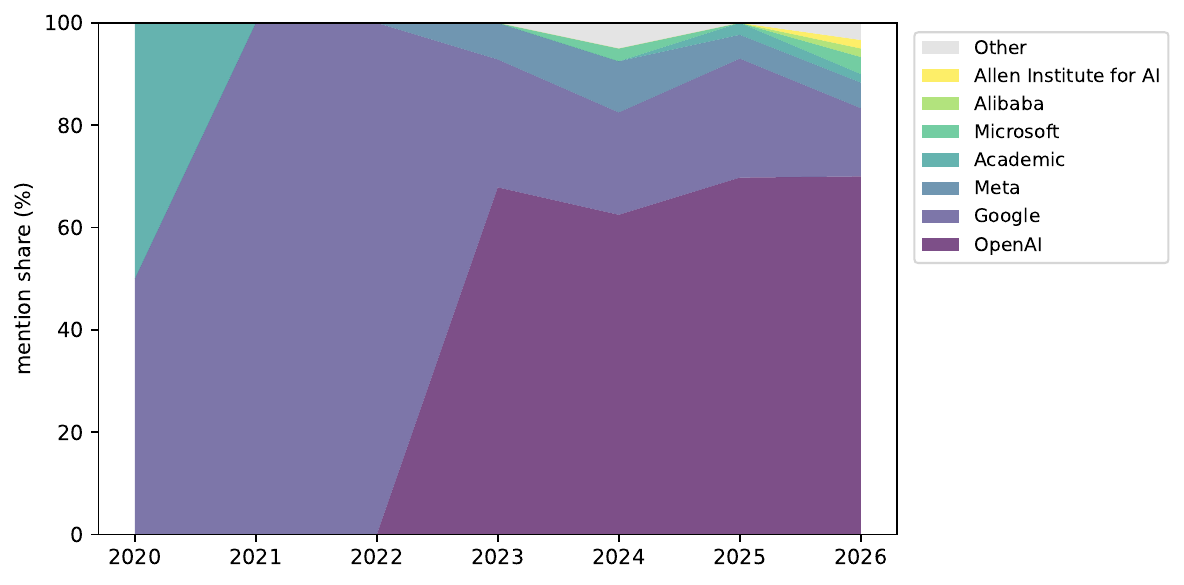}}
{LLM provider market shares.\label{fig:provider_share}}
{We chronologically represent the share of models listed in titles and abstracts grouped by the organizations that develop them. In 2020, the lead from Google is due to the BERT model. Subsequently, academics have retrained BERT on domain-specific data, leading to specialized BERT versions, such as FinBERT, BioBERT, ClinicalBERT, Legal-BERT. These models fall under the umbrella of the "Academic" category in the legend.}
\end{figure}

\section{Temperature timeline}\label{sec:temperature}

\begin{figure}[!h]
\FIGURE
{\includegraphics[width=\textwidth]{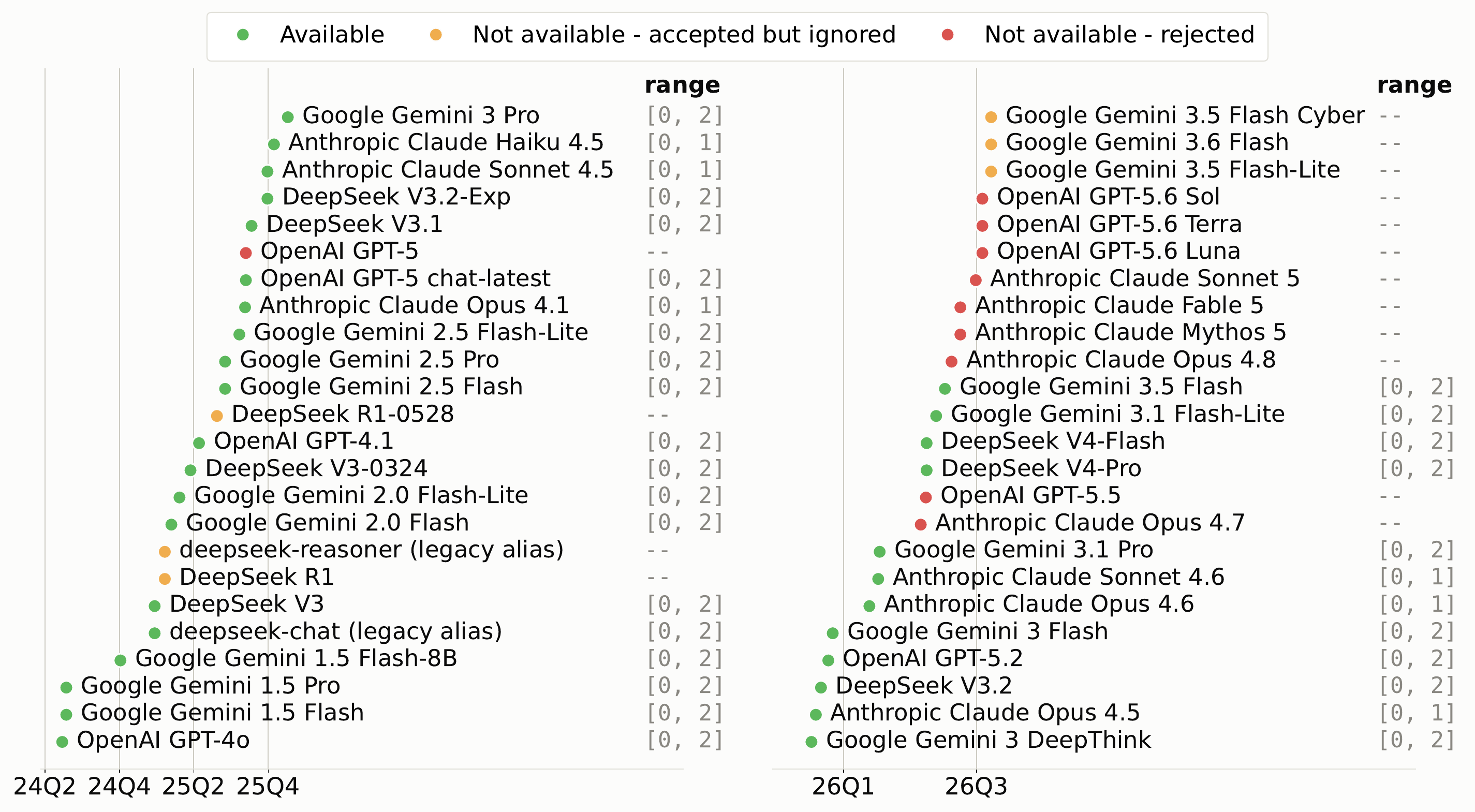}}
{LLM temperature timeline (mid-2024 to mid-2026).\label{fig:temperature}}
{We chronologically represent the release of models from Anthropic, DeepSeek, Google and OpenAI. Each model is associated with a policy regarding temperature: whether it can be user-specified (green), or whether it is simply ignored (orange), or prohibited (red). When temperature can be set, the range columns provide the range of possible values. The information was gathered from the providers' websites directly (for 75\% of points), and, when not available, from API calls, or from Wikipedia.}
\end{figure}
\fi

\end{APPENDICES}

\ACKNOWLEDGMENT{We thank conference participants at the AI in Economics and Finance Conference at Uni Turin for helpful comments. The replication package of this paper is available at \url{https://doi.org/10.5281/zenodo.24198290}.}




\renewcommand{\bibfont}{\small\renewcommand{\baselinestretch}{1.0}\selectfont} 
\setlength{\bibsep}{6pt}

\ifarxiv
  \bibliographystyle{plainnat}
\else
  \bibliographystyle{informs2014}
\fi

\bibliography{bib}

@article{bail2024,
  author  = {Bail, C. A.},
  title   = {Can generative {AI} improve social science?},
  journal = {Proceedings of the National Academy of Sciences},
  year    = {2024},
  volume  = {121},
  number  = {21},
  pages   = {e2314021121}
}

@article{charness2023,
  author      = {Charness, G. and Jabarian, B. and List, J. A.},
  title       = {Generation next: {E}xperimentation with {AI}},
  institution = {National Bureau of Economic Research},
  year        = {2023},
  journal      = {NBER Working Paper},
  number      = {31679}
}

@article{chen2024,
  author  = {Chen, L. and Zaharia, M. and Zou, J.},
  title   = {How is {ChatGPT's} behavior changing over time?},
  journal = {Harvard Data Science Review},
  year    = {2024},
  volume  = {6},
  number  = {2}
}

@article{christensen2018,
  author  = {Christensen, G. and Miguel, E.},
  title   = {Transparency, reproducibility, and the credibility of economics research},
  journal = {Journal of Economic Literature},
  year    = {2018},
  volume  = {56},
  number  = {3},
  pages   = {920 to 980}
}

@article{dell2025,
  author  = {Dell, M.},
  title   = {Deep learning for economists},
  journal = {Journal of Economic Literature},
  year    = {2025},
  volume  = {63},
  number  = {1},
  pages   = {5 to 58}
}

@article{fivsar2024reproducibility,
  title={Reproducibility in {Management Science}},
  author={Fi{\v{s}}ar, Milo{\v{s}} and Greiner, Ben and Huber, Christoph and Katok, Elena and Ozkes, Ali I and {Management Science Reproducibility Collaboration}},
  journal={Management Science},
  volume={70},
  number={3},
  pages={1343--1356},
  year={2024},
  publisher={Informs}
}

@article{gilardi2023,
  author  = {Gilardi, F. and Alizadeh, M. and Kubli, M.},
  title   = {{ChatGPT} outperforms crowd workers for text annotation tasks},
  journal = {Proceedings of the National Academy of Sciences},
  year    = {2023},
  volume  = {120},
  number  = {30},
  pages   = {e2305016120}
}

@article{hansen2023,
  author = {Hansen, A. L. and Kazinnik, S.},
  title  = {Can {ChatGPT} decipher {Fedspeak}?},
  year   = {2024},
  journal   = {SSRN Working Paper},
  volume= {4399406}
}

@article{harford2026,
  author    = {Harford, Jarrad and He, Qiyang and Qiu, Buhui},
  title     = {Firm-Level Labor Shortage Exposure},
  journal   = {Review of Financial Studies},
  volume    = {Forthcoming},
  year      = {2026}
}

@article{he2025,
  author  = {He, H. and {Thinking Machines Lab}},
  title   = {Defeating nondeterminism in {LLM} inference},
  journal = {Thinking Machines Lab: Connectionism},
  year    = {2025},
  doi     = {10.64434/tml.20250910}
}

@article{horton2023,
  author      = {Horton, J. J. and Filippas, A. and Manning, B. S.},
  title       = {Large language models as simulated economic agents: {W}hat can we learn from homo silicus?},
  year        = {2026},
  journal = {arXiv Prepring},
  number        = {2301.07543}
}

@article{kim2026,
  author = {Kim, A. G.},
  title  = {Tractable use of large language models in financial economics research},
  year   = {2026},
  journal   = {Working paper, University of Chicago}
}

@misc{koren2022,
  author       = {Koren, M. and Connolly, M. and Llull, J. and Vilhuber, L.},
  title        = {Data and code availability standard},
  year         = {2022},
  note         = {Version 1.0},
  howpublished = {Zenodo}
}

@article{korinek2023,
  author  = {Korinek, A.},
  title   = {Generative {AI} for economic research: {U}se cases and implications for economists},
  journal = {Journal of Economic Literature},
  year    = {2023},
  volume  = {61},
  number  = {4},
  pages   = {1281 to 1317}
}

@techreport{korinek2025,
  author      = {Korinek, A.},
  title       = {{AI} agents for economic research},
  institution = {National Bureau of Economic Research},
  year        = {2025},
  journal        = {NBER Working Paper},
  number      = {34202}
}

@inproceedings{kweon2025uncertainty,
  title={Uncertainty quantification and decomposition for {LLM}-based recommendation},
  author={Kweon, Wonbin and Jang, Sanghwan and Kang, SeongKu and Yu, Hwanjo},
  booktitle={Proceedings of the ACM on Web Conference 2025},
  pages={4889--4901},
  year={2025}
}

@article{loch2025reinforcing,
  title={Reinforcing Research Transparency at {Management Science}},
  author={Loch, Christoph},
  journal={Management Science},
  volume={71},
  number={9},
  pages={7--8},
  year={2025},
  publisher={INFORMS}
}

@article{lopezlira2025,
  author  = {{Lopez Lira}, A. and Tang, Y.},
  title   = {Can {ChatGPT} forecast stock price movements? {R}eturn predictability and large language models},
  journal = {Journal of Financial Economics},
  year    = {2025},
  note    = {Forthcoming}
}

@article{loughran2011liability,
  title={When is a liability not a liability? {T}extual analysis, dictionaries, and {10-Ks}},
  author={Loughran, Tim and McDonald, Bill},
  journal={Journal of Finance},
  volume={66},
  number={1},
  pages={35--65},
  year={2011},
  publisher={Wiley Online Library}
}

@article{ollion2024,
  author  = {Ollion, {\'E}. and Shen, R. and Macanovic, A. and Chatelain, A.},
  title   = {The dangers of using proprietary {LLMs} for research},
  journal = {Nature Machine Intelligence},
  year    = {2024},
  volume  = {6},
  number  = {1},
  pages   = {4 to 5}
}

@article{ouyang2024,
  author  = {Ouyang, S. and Zhang, J. M. and Harman, M. and Wang, M.},
  title   = {An empirical study of the non-determinism of {ChatGPT} in code generation},
  journal = {ACM Transactions on Software Engineering and Methodology},
  year    = {2024},
  volume  = {34},
  number  = {2},
  pages   = {42},
  doi     = {10.1145/3697010}
}

@article{palmer2024,
  author  = {Palmer, A. and Smith, N. A. and Spirling, A.},
  title   = {Using proprietary language models in academic research requires explicit justification},
  journal = {Nature Computational Science},
  year    = {2024},
  volume  = {4},
  number  = {1},
  pages   = {2 to 3}
}

@article{perignon2024a,
  author  = {P{\'e}rignon, C.},
  title   = {The role of third party verification in research reproducibility},
  journal = {Harvard Data Science Review},
  year    = {2024},
  volume  = {6.2}
}

@article{perignon2019,
  author  = {P{\'e}rignon, C. and Gadouche, K. and Hurlin, C. and Silberman, R. and Debonnel, E.},
  title   = {Certify reproducibility with confidential data},
  journal = {Science},
  year    = {2019},
  volume  = {365},
  number  = {6449},
  pages   = {127 to 128}
}

@article{perignon2024b,
  author  = {P{\'e}rignon, C. and Akmansoy, O. and Hurlin, C. and Dreber, A. and Holzmeister, F. and Huber, J. and Johannesson, M. and Kirchler, M. and Menkveld, A. J. and Razen, M. and Weitzel, U.},
  title   = {Computational reproducibility in finance: {E}vidence from 1,000 tests},
  journal = {Review of Financial Studies},
  year    = {2024},
  volume  = {37},
  number  = {11},
  pages   = {3558 to 3593}
}

@article{siano2025news,
  title={The news in earnings announcement disclosures: Capturing word context using {LLM} methods},
  author={Siano, Federico},
  journal={Management Science},
  volume={71},
  number={11},
  pages={9831--9855},
  year={2025},
  publisher={INFORMS}
}

@article{spirling2023,
  author  = {Spirling, A.},
  title   = {Why open source generative {AI} models are an ethical way forward for science},
  journal = {Nature},
  year    = {2023},
  volume  = {616},
  number  = {7957},
  pages   = {413}
}

@article{tetlock2008more,
  title={More than words: Quantifying language to measure firms' fundamentals},
  author={Tetlock, Paul C and Saar-Tsechansky, Maytal and Macskassy, Sofus},
  journal={Journal of Finance},
  volume={63},
  number={3},
  pages={1437--1467},
  year={2008},
  publisher={Wiley Online Library}
}

@article{vilhuber2019,
  author  = {Vilhuber, L.},
  title   = {Report by the {AEA} data editor},
  journal = {AEA Papers and Proceedings},
  year    = {2019},
  volume  = {109},
  pages   = {718 to 729}
}

@article{vilhuber2020,
  author  = {Vilhuber, L.},
  title   = {Reproducibility and replicability in economics},
  journal = {Harvard Data Science Review},
  year    = {2020},
  volume  = {2},
  number  = {4}
}

@misc{vilhuber2022,
  author       = {Vilhuber, L. and Connolly, M. and Koren, M. and Llull, J. and Morrow, P.},
  title        = {A template {README} for social science replication packages},
  year         = {2022},
  note         = {Version 1.1},
  howpublished = {Social Science Data Editors, Zenodo}
}

@article{wang2025assessing,
  title={Assessing consistency and reproducibility in the outputs of large language models: {E}vidence across diverse finance and accounting tasks},
  author={Wang, Julian Junyan and Wang, Victor Xiaoqi},
  journal={arXiv Preprint},
  number={2503.16974},
  year={2025}
}

@article{hayes2024,
  author       = {Hayes, Jamie and Shumailov, Ilia and Yona, Itay},
  title        = {Buffer Overflow in Mixture of Experts},
  journal      = {arXiv Preprint},
  number       = {2402.05526},
  year         = {2024}
}

@article{fedus2022switch,
  author  = {Fedus, William and Zoph, Barret and Shazeer, Noam},
  title   = {Switch Transformers: Scaling to Trillion Parameter Models with Simple and Efficient Sparsity},
  journal = {Journal of Machine Learning Research},
  year    = {2022},
  volume  = {23},
  number  = {120},
  pages   = {1--39},
  url     = {https://jmlr.org/papers/v23/21-0998.html}
}

@article{deKokMS2025,
author = {de Kok, Ties},
title = {Chat{GPT} for Textual Analysis? {H}ow to Use Generative {LLMs} in Accounting Research},
journal = {Management Science},
volume = {71},
number = {9},
pages = {7888-7906},
year = {2025}
}

@article{feuerriegel2026,
  author  = {Feuerriegel, S. and Barrie, C. and Crockett, M. J. and Globig, L. and McLoughlin, K. and others},
  title   = {A reporting checklist for large language models in behavioural science},
  journal = {Nature Human Behaviour},
  year    = {2026},
  volume  = {10},
  number  = {7},
  pages   = {1182 to 1186},
  doi     = {10.1038/s41562-026-02492-7}
}

@misc{higton2026,
  author       = {Higton, J. and Tucker, J. A. and Barrie, C.},
  title        = {Request routing services undermine the reproducibility of open weight {LLMs}},
  year         = {2026},
  howpublished = {SocArXiv preprint},
  doi          = {10.31235/osf.io/sr9hy_v1}
}

@misc{anthropic2026fable5,
  author       = {{Anthropic}},
  title        = {Redeploying {Claude} {Fable} 5},
  year         = {2026},
  howpublished = {\url{https://www.anthropic.com/news/redeploying-fable-5}},
  note         = {Published June 30, 2026, updated July 1, 2026. Accessed November 6, 2026}
}

@misc{tgi2026,
  author       = {{Hugging Face}},
  title        = {Text Generation Inference: validation.rs},
  year         = {2026},
  howpublished = {\url{https://github.com/huggingface/text-generation-inference/blob/main/router/src/validation.rs}},
  note         = {Accessed November 4, 2026}
}

@misc{vllm2026,
  author       = {{vLLM}},
  title        = {Sampling parameters},
  year         = {2026},
  howpublished = {\url{https://docs.vllm.ai/en/latest/api/vllm/sampling_params/}},
  note         = {Accessed November 4, 2026}
}

@misc{mlxvlm2026,
  author       = {{MLX-VLM}},
  title        = {sample\_utils.py},
  year         = {2026},
  howpublished = {\url{https://github.com/Blaizzy/mlx-vlm/blob/main/mlx_vlm/sample_utils.py}},
  note         = {Accessed November 4, 2026}
}

@misc{openai2026gpt6,
  author       = {{OpenAI}},
  title        = {Using {GPT-6}},
  year         = {2026},
  howpublished = {\url{https://developers.openai.com/api/docs/guides/latest-model}},
  note         = {Accessed November 4, 2026}
}

@misc{openai2026responses,
  author       = {{OpenAI}},
  title        = {Create a model response},
  year         = {2026},
  howpublished = {\url{https://developers.openai.com/api/reference/resources/responses/methods/create}},
  note         = {Accessed November 4, 2026}
}

@misc{deepseek2026api,
  author       = {{DeepSeek}},
  title        = {Create chat completion},
  year         = {2026},
  howpublished = {\url{https://api-docs.deepseek.com/api/create-chat-completion}},
  note         = {Accessed November 4, 2026}
}

@misc{anthropic2026opus55,
  author       = {{Anthropic}},
  title        = {Migrating to {Claude} {Opus} 5.5},
  year         = {2026},
  howpublished = {\url{https://platform.claude.com/docs/en/models/opus-5-5/migration-guide}},
  note         = {Accessed November 4, 2026}
}

@misc{deepseek2026thinking,
  author       = {{DeepSeek}},
  title        = {Thinking mode},
  year         = {2026},
  howpublished = {\url{https://api-docs.deepseek.com/guides/thinking_mode/}},
  note         = {Accessed November 4, 2026}
}

@misc{google2026prompting,
  author       = {{Google}},
  title        = {Prompting strategies},
  year         = {2026},
  howpublished = {\url{https://aistudio.google.com/docs/prompting-strategies}},
  note         = {Accessed November 4, 2026}
}

@misc{mnsc2026policy,
  author       = {{Management Science}},
  title        = {Code and data disclosure policy},
  year         = {2026},
  howpublished = {\url{https://pubsonline.informs.org/page/mnsc/code-and-data-disclosure-policy}},
  note         = {Effective June 1, 2019; revised April 20, 2026. Accessed November 6, 2026}
}

@misc{mnsc2024checklist,
  author       = {{Management Science}},
  title        = {Code and data replication package checklist},
  year         = {2024},
  note         = {Version 5.16.2024}
}

@misc{opre2026policy,
  author       = {{Operations Research}},
  title        = {{OPRE} code and data disclosure policy},
  year         = {2026},
  howpublished = {\url{https://pubsonline.informs.org/page/opre/code-and-data-disclosure-policy}},
  note         = {Accessed November 19, 2026}
}

@misc{opre2026instructions,
  author       = {{Operations Research}},
  title        = {Procedures for submission of accompanying software and data},
  year         = {2026},
  howpublished = {\url{https://orjournal.github.io/InstructionsForAuthors.html}},
  note         = {Accessed November 19, 2026}
}

@misc{msom2026ethics,
  author       = {{Manufacturing \& Service Operations Management}},
  title        = {Guidelines for ethical behavior in publishing},
  year         = {2026},
  howpublished = {\url{https://pubsonline.informs.org/page/msom/editorial-mission}},
  note         = {Accessed November 19, 2026}
}

@misc{isre2026ethics,
  author       = {{Information Systems Research}},
  title        = {Guidelines for ethical behavior in publishing},
  year         = {2026},
  howpublished = {\url{https://pubsonline.informs.org/page/isre/guidelines-for-ethical-behavior-in-publishing}},
  note         = {Accessed November 19, 2026}
}

@article{desai2013,
  author  = {Desai, Preyas S.},
  title   = {Editorial---{Marketing Science} replication and disclosure policy},
  journal = {Marketing Science},
  year    = {2013},
  volume  = {32},
  number  = {1},
  pages   = {1--3},
  doi     = {10.1287/mksc.1120.0761}
}


\end{document}